# Do more heads imply better performance? An empirical study of team thought leaders' impact on scientific team performance

**Yi Zhao[1], Yuzhuo Wang[2], Heng Zhang[1], Donghun Kim[3], Chao Lu[4], Yongjun Zhu[3*], Chengzhi Zhang[1*]**

1. Department of Information Management, Nanjing University of Science and Technology, Nanjing, China {yizhao93, zh_heng, zhangcz}@njust.edu.cn
2. School of Management, Anhui University, Hefei, China wangyuzhuo@ahu.edu.cn
3. Department of Library and Information Science, Yonsei University, Seoul, Republic of Korea {dhkim91, zhu}@yonsei.ac.kr
4. Business School, Hohai University, Nanjing, China luchao91@hhu.edu.cn

**Abstract**

Thought leadership plays a crucial role in boosting team performance; thus, teams with more thought leaders may perform better. However, the impact of the number of thought leaders on team performance in a scientific context remains understudied. In this study, we consider the authors of a publication as a scientific team and define the authors responsible for conceptual tasks, i.e., "conceived and designed the experiments" (one of the tasks described in the PLOS contribution statements classification system), as thought leaders. Leveraging more than 140,000 papers from PLOS journals, we examine the relationship between the number of thought leaders and two aspects of team performance (i.e., team impact and team disruptiveness) from both correlational and causal perspectives. The results showed that (1) an inverted U-shaped relationship exists between the number of thought leaders and the team's impact, and (2) teams with more thought leaders tend to produce less disruptive ideas. We also explored the impact of international collaboration, team size, and gender diversity together with the number of thought leaders on team performance and found that (3) international collaboration improves team impact but lowers the disruptiveness of team outputs. This study advances scholarly understanding of thought leadership in scientific teams and provides valuable insights for policymakers and team managers.

---

[*] Corresponding author, Email: zhu@yonsei.ac.kr (Yongjun Zhu), zhangcz@njust.edu.cn (Chengzhi Zhang)

## 1. Introduction

As scientific research becomes increasingly complex, scientific innovation is ever more a team endeavor (Wuchty et al. 2007). Conceiving new ideas is the foundation of scientific innovation and the most important step in conducting research. Our study refers to members of a scientific team who contribute to generating research ideas as thought leaders. This approach follows the work of McCrimmon (2005), which defined thought leaders as people who are responsible for proposing and championing new ideas rather than for managing a team. Based on the author contribution statements, Xu, Wu, et al. (2022) define leadership roles within scientific teams as those researchers engaged in leading contributions, including activities with keywords such as "design," "supervise," and "write." Seniority is also a prevalent manifestation of leadership. Xu et al. (2024) argue that authors whose academic careers have lasted more than ten years often take on leadership roles within scientific teams. However, leadership discussed in these studies generally has managerial authority and is not a pure form of thought leadership. To our knowledge, there has been a lack of research exploring the role of thought leaders within scientific teams. Thought leaders play a vital role in enhancing breakthrough innovation in scientific research (Harvey et al. 2021), which has prompted us to deepen our understanding of thought leaders in a scientific context.

In response to the dramatic growth of human knowledge, scholars are compelled to narrow their research foci to quickly reach the frontiers of knowledge, which inevitably enables them to become more specialized (Jones 2009). In the era of big science, creating useful and novel ideas requires the integration of diverse individual expertise (Bercovitz & Feldman 2011). As a well-known proverb says, "Two heads are better than one," suggesting that teams with more thought leaders may have greater potential to produce impactful and novel outputs. In experimental studies, scholars indicated that collaborative groups perform better than individuals on various tasks, including perceptual tasks (Koriat 2012), letters-to-numbers tasks (Laughlin et al. 2006), and so on. However, several studies have found conflicting results. Drawing on the data from Wikipedia articles, Greenstein and Zhu (2016) reported that articles with many contributing authors are more neutral. An experimental study also revealed that groups tend to reject external information in decision-making more than individuals (Minson & Mueller 2012). In summary, the conclusions of these studies are inconsistent. It remains unclear whether "more heads" typically result in better performance in scientific teams.

Scientific research is regarded as "knowledge-intensive" work, primarily of an intellectual nature (Wagner et al. 2011). In most research cases, team members collaborate, exchange expertise and resources, collectively brainstorm possible solutions to scientific problems, and then produce scientific outputs. Intuitively, involving more team members in the conception of ideas (i.e., as thought leaders) can enhance the quality of research outputs. However, no evidence supports the assertion that a large number of thought leaders benefit scientific team performance. Understanding how the number of thought leaders within teams affects team performance is of utmost importance for team managers and policymakers. In the field of science of science, citation counts are commonly used to measure team impact (Yang, Tian, et al. 2022), whereas the disruption index measures the degree to which a scientific idea disrupts previous ideas (Funk & Owen-Smith 2017). Both measures were used in this study to examine the effect of the number of thought leaders on team performance. To date, the correlation between the number of thought leaders and team performance remains uncertain. Furthermore, the potential causal effect of the number of thought leaders on team performance is still unexplored. To close these research gaps, the following questions are proposed:

***RQ1***: What is the relationship between the number of thought leaders and team performance?

***RQ2***: How do we investigate a causal relationship between the number of thought leaders and team performance?

The main contribution of this paper is three-fold. Firstly, we extend the concept of thought leader from organizational management to the context of scientific innovation, which advances our understanding of thought leadership in scientific teams. Secondly, we explore the association between the number of thought leaders and two aspects of team performance: team impact and disruptiveness. Thirdly, we employ a novel matching strategy to examine the potential causal effect of the number of thought leaders, going beyond mere estimation of the correlation between the number of thought leaders and team performance.

## 2. Literature Review

Leadership plays a crucial role in fostering team learning and innovation (Klaic et al. 2020). In the context of scientific innovation, the quality of research ideas determines the value and impact of scientific discoveries (Ke 2020). However, leaders within scientific teams are not always directly engaged in generating ideas (Larivière et al. 2016; Xu, Wu, et al. 2022). Therefore, to accurately investigate the association between leadership and scientific team performance, it is necessary to consider the specific roles and responsibilities of leaders. More specifically, we should examine the effect of leaders who are tasked with conceiving ideas on team performance. The concept of thought leadership was first proposed by McCrimmon (2005), and it has not been studied in the science of science. Therefore, we review studies exploring 1) general leadership in scientific teams, 2) the order of authors in association with leadership, and 3) the relationship between the team size and team performance.

### 2.1 Leadership in scientific teams

Some studies distinguish leading and supporting roles in teams based on the specific tasks assigned to individual team members. For example, Robinson-Garcia et al. (2020) utilized archetypal analysis to identify leader archetypes based on their contributorships. The authors found that the leaders exhibited a high involvement in various research tasks, especially in “wrote the manuscript” and “conceived and designed the experiments.” Similarly, Xu, Wu, et al. (2022) defined team leaders as individuals engaged in specific tasks, including “conceive,” “design,” “lead,” “supervise,” “coordinate,” “interpret,” and “write.” In these studies, the authors believed that multiple leaders share the leading position and manage scientific teams. This type of leadership is referred to as shared leadership in the field of management (Carson et al. 2007), which is well-suited for the current team-based research environment. Some scholars argued that adopting shared leadership can enhance team performance (Carson et al. 2007; D'Innocenzo et al. 2016; Perry et al. 1999; Klavans & Boyack 2010). Shared leadership involves managing people, making decisions, and coordinating teams. Different from shared leadership, thought leadership aims to propose and promote new ideas. Scholars believe that thought leadership is more important than shared leadership when innovation is crucial to the organization's development and success (McCrimmon 2005).

### 2.2 The order of authors and leadership

The order of authors in an article byline is a useful clue for identifying leadership (Chinchilla-Rodriguez et al. 2019). Despite disciplinary differences in authorship norms, studies reported that the first and last authors are more significant contributors than the middle (but not corresponding) authors in publications (Rennie et al. 2000; Frandsen & Nicolaisen 2010). In most cases, the first and last positions in article bylines are used as proxies for leadership roles (van Leeuwen 2008), because these authors, especially the last author, usually contribute to conceiving ideas (Larivière et al. 2016). However, a recent study indicated that middle authors are increasingly responsible for the task of generating ideas (Sauermann & Haeussler 2017). Hence, ignoring the contribution of idea conceptualization by middle authors may bias our understanding of the division of labor within teams and team formation. In this study, we expand the traditional approach of considering the first and the last authors as thought leaders and additionally explore middle authors as potential thought leaders.

### 2.3 Team size and team performance

Due to the lack of studies examining the relationship between the number of thought leaders and scientific team performance, we review studies on the relationship between team size and team performance outside and within the science of science field.

In the setting of creative use tasks, knowledge diversity among team members benefits the creative outcome, whereas superficial diversity, such as gender and race, does not affect the creative outcome (Dutcher & Rodet 2022). In solving letters-to-numbers problems, Laughlin et al. (2006) indicated that an increased group size enhances problem-solving performance, especially for highly collective problems. However, Koriat (2012) suggested that collaborative groups do not always outperform independent individuals in perceptual tasks, depending on the specific situation and team members' confidence. Swaab et al. (2014) discovered an inverted U-shaped relationship between talent size and sports team performance. This relation can be moderated by the level of task interdependence. Using panel data from the National Basketball Association teams, Szatmari (2022) confirmed that an increase in talent amount improves team performance, but only to a certain extent. Beyond that, the author also found the moderating effect of talents' age diversity. Although the conclusions are drawn from sports teams rather than knowledge-intensive teams, the findings shed light on the relationship between talent size and team performance.

In the science of science, team performance is defined by the extent to which a team achieves its scientific goals, along with the quality and quantity of its scientific outcomes (Bertolotti et al. 2015). Previous studies adopted indicators that measure the quality and impact of scientific outputs produced by research teams as proxies for team performance, including citation counts, novelty (Lee et al. 2015), disruptiveness (Funk & Owen-Smith 2017), and hit paper status (Yang, Tian, et al. 2022). Larivière et al. (2015) used three indicators as proxies for team collaboration: the number of authors, their addresses, and countries of origin. The results show that larger teams tend to receive more citations. Using millions of article teams, patent teams, and software development teams, Wu et al. (2019) indicated that small teams tend to generate disruptive ideas, while large teams are more likely to produce incremental innovations. They used the number of authors as a proxy for team size to explore the impact of team size on the disruptiveness of ideas; however, not all the authors are involved in conceiving ideas and serve as thought leaders in teams. Thus, the relationship between the number of thought leaders and team performance remains unknown, and this study aims to fill this research gap.

## 3. Data and Methods

### 3.1 Data collection

In this study, we collected 162,613 research articles published in the Public Library of Science[1] (PLOS) from 2007 to 2015 in XML format. PLOS requires authors to disclose their contributions to the research they report. Scholars have used author contribution statements in PLOS articles to conduct studies on author credit allocation (Yang, Xiao, et al. 2022), patterns of scientific collaboration (Xu, Wu, et al. 2022; Lu et al. 2020), and the relationship between contribution and byline orders (Lu et al. 2022; Corrêa Jr et al. 2017; Sauermann & Haeussler 2017). The classification systems for contribution statements before and after 2006 are inconsistent, so we restricted the sample to articles published from 2007–2015.

As shown in Figure 1, five tasks are commonly used in PLOS articles (Larivière et al. 2016): "Conceived and designed the experiments," "Performed the experiments," "Analyzed the data," "Contributed reagents/materials/analysis tools" and "Wrote the paper." In our study, the authors who perform the task of "conceived and designed the experiments" are regarded as thought leaders in teams. Following the practice of Lu et al. (2022), we first parsed author contribution statements from full-text articles in XML format and converted them into pairs of author name acronyms and tasks. Second, we converted authors' full names from the authorship list into a standardized format of initials, retaining only the initial letter of their first, middle (if present) and last names. This step creates a mapping table between authors' full names and their acronyms. Third, we linked the authors' full names and tasks based on the authors name acronyms using string matching (See Fig.1). For the few unmatched cases, we used the Levenshtein distance algorithm (Levenshtein 1966) to compute the similarity between authors' full names in the byline and their acronyms disclosed in the author contribution statement. Then, we mapped the most similar ones. To verify the performance of the matching approach, we randomly selected a sample of 100 papers, which included 1,612 pairs of author full names and their associated tasks. Each pair was manually reviewed to ascertain the correctness of the matches. We found an overall accuracy of 99.19% in the matching results. Articles in which all authors' contribution statements are fully extracted were preserved, and articles failing to successfully parse the “conceived and designed the experiments” contribution type were also excluded from the dataset. We also removed the articles in which every author participated in every type of contribution, leaving 151,803 articles.

[1] https://plos.org/

RESEARCH ARTICLE

Nocturnal Homing: Learning Walks in a Wandering Spider?

Thomas Nørgaard, Yakir L. Gagnon, Eric J. Warrant

Published: November 7, 2012 • https://doi.org/10.1371/journal.pone.0049263

<fn fn-type="con">
<p>Conceived and designed the experiments: TN. Performed the experiments: TN. Analyzed the data: TN YLG. Contributed reagents/materials/analysis tools: TN YLG EJW. Wrote the paper: TN YLG EJW.</p>
</fn>

Figure 1 Matching authors’ full names with their initials

Articles’ citing relations and disambiguated author names are essential metadata for this study. PLOS does not provide this information, and we linked the PLOS dataset with PubMed Knowledge Graph (PKG) (Xu et al. 2020). PKG provides information relevant to publications, such as citation relationships, keywords, disambiguated author names, etc. PKG integrated the disambiguation results from both Author-ity and Semantic Scholar. As a result, the $F_1$ score of author name disambiguation (AND) achieved 98.09%. This dataset has been used in several studies (Huo et al. 2022; Sourati & Evans 2023; Zhao et al. 2022). The PLOS publisher provides the digital object identifier (DOI) to index and track the articles, while PMID (i.e., PubMed reference number) is commonly used as the unique identifier for each article in the PKG. The linking process consists of two steps. First, we obtained each PLOS article’s PMID via the ID Converter[2] provided by the National Library of Medicine. Second, we linked the PLOS dataset and PKG based on the PMID of the articles. Subsequently, we retained articles in which all authors had unique identifiers and no less than five references. After this process, our dataset had 145,800 research articles by 594,049 scholars.

## 3.2 Methods

### 3.2.1 Data processing

#### (1) Name-to-gender inference

NamSor[3] was used to infer the gender of authors. NamSor, a paid commercial service, used a Naive Bayes model for gender designation, which was trained on 1.3 million people’s names from around the world. Additionally, NamSor is one of the very few gender-inference services that can leverage first and last names for gender designation. According to a recent study, NamSor has achieved the best performance with around 2% misclassifications among commonly used name-to-gender services, including Gender API, genderize.io, and Wiki-Gendersort (Sebo 2021). NamSor's application for inferring author genders from names is notably prevalent in science of science (Yang, Tian, et al. 2022; Zhang et al. 2022). To evaluate the performance of NamSor for our data, we compared the deduced gender labels of authors against a ground truth dataset composed of 2,000 male and female full names manually compiled by Larivière et al. (2013). Out of the 384 author names that coincided with our data, 337 were consistent with the gender labels of the ground truth dataset, yielding an accuracy of 87.76%. Subsequently, we only retained articles with at least two authors, as we defined teams to have a minimum of two members. After this step, we obtained a dataset of 341,925 male and 252,124 female scholars from 145,800 research articles.

[2] https://www.ncbi.nlm.nih.gov/pmc/tools/idconv/
[3] https://namsor.app/

**(2) Discipline identification**

We also identified the discipline of each PLOS article. According to the practice of Lu et al. (2019), 130,929 (90%) of PLOS articles were linked to Web of Science (WoS) records, and disciplines were determined based on the National Science Foundation disciplinary classification system. The disciplinary distribution of our dataset is shown in Table 1.

Table 1 The disciplinary distribution of the dataset

| Discipline | # of publications | Ratio (%) |
|---|---|---|
| Clinical Medicine | 60,777 | 41.69 |
| Biomedical Research | 48,209 | 33.07 |
| Unmatched | 14,871 | 10.19 |
| Biology | 11,599 | 7.96 |
| Mathematics | 2,803 | 1.92 |
| Psychology | 2,668 | 1.83 |
| Earth and Space | 1,142 | 0.78 |
| Health | 1,050 | 0.72 |
| Physics | 710 | 0.49 |
| Engineering and Technology | 590 | 0.40 |
| Social Sciences | 570 | 0.39 |
| Chemistry | 550 | 0.38 |
| Professional Fields | 253 | 0.17 |
| Humanities | 8 | 0.01 |
| Total | 145,800 | 100 |

**Note**: The term “Unmatched” in the first columns indicates articles that failed to link to WoS records.

**(3) Institution and country identification**

PLOS does not provide disambiguated author affiliation information. We matched PLOS articles to OpenAlex[4] records based on the DOI of each article. OpenAlex is a reputable and free bibliographic database that provides disambiguated authors' affiliation and country information. After the linking, we could accurately obtain the affiliations and country metadata of the authors in each article. We only retained articles in which all author's affiliations and country information was identifiable. The final dataset used for further analysis included 141,550 articles.

### 3.2.2 Dependent, independent, and control variables

**(1) Dependent variables**

**Team performance.** Citation counts have been widely used to measure team impact in the science of science (Shen et al. 2022), although they cannot adequately capture the research value of scientific articles (Sugimoto 2021). Recently, a metric called the disruption index was used to measure the degree to which a scientific idea disrupts previous ideas (Wu et al. 2019), which originated from Funk and Owen-Smith (2017)'s CD index. In this study, we used *team impact* and *disruptiveness* to measure team performance. Citation counts were employed as a proxy for team impact, while the disruption index serves as a proxy for team disruptiveness.

To obtain citation counts, we followed the practice of previous studies on team performance, where a five-year cumulative citation count was used (Wang 2012; Liu et al. 2022). A detailed explanation of obtaining annual citations can be found in Section 3.1. For the disruption index (Wu et al. 2019), the calculation method is as follows:

$$Disruption\ index = \frac{n_i - n_j}{n_i + n_j + n_k} \quad (1)$$

Where $n_i$ represents backward citations that cite the focal paper only, $n_j$ indicates backward citations that cite the focal paper and parts of its references simultaneously, and $n_k$ represents backward citations citing references of the focal paper only. We transformed the disruption index into the disruption percentile for the analysis, following Wu et al. (2019).

**(2) Independent variables**

**Ratio of thought leaders.** In this study, we treated the authors who "conceived and designed the experiments" as thought leaders in teams. Referring to Swaab et al. (2014), we created a team-level measure by counting the number of thought leaders in each team and dividing this number by the size of the team, i.e., the ratio of thought leaders in a team.

**(3) Control variables**

We included several variables to control for team characteristics (i.e., team size, institutional diversity, international team, number of references, gender diversity, year dummies, and field dummies) and thought leaders (i.e., average academic age, average prior productivity).

**Team size.** Previous studies have argued that team size affects team impact (Wuchty et al. 2007) and the disruptiveness of team outcomes (Wu et al. 2019). Therefore, we included this variable in our models and used the number of authors in an article as a proxy of team size.

---

[4] https://docs.openalex.org/

**Institutional diversity.** We counted the number of distinct institutions in an article to proxy a team's institutional diversity. For example, paper *P* has three authors, affiliated with institution *A*, institution *B*, and institution *A*, respectively. Then, the institutional diversity of paper *P* is 2. A team with members from multiple institutions was found to perform better in producing innovative ideas (AlShebli et al. 2018).

**International team.** We defined team internationality as a binary variable represented by 1. An international team's members are from at least two different countries. Scientific outputs produced by international teams are more likely to receive more citations than those produced by domestic teams (Wagner et al. 2019).

**Number of references.** The number of references in the focal paper affects the calculation of the disruption index (Ruan et al. 2021; Lee et al. 2015). Thus, we included the number of references as a control variable.

**Gender diversity.** Similar to the international team variable, we adopted a binary method to measure team gender diversity. We used 1 to denote a team composed of men and women, called a mixed-gender team, and 0 for a same-gender team. Previous studies have demonstrated that gender diversity influences team creativity (Yang, Tian, et al. 2022; Lee & Chung 2022).

**Year and field dummies.** Previous studies have observed that differences in citation count and disruption exist across years and fields (Park et al. 2023; Wang et al. 2013; Thelwall & Fairclough 2015). Therefore, our subsequent analysis included year dummies and field dummies.

**Average academic age of thought leader(s).** Academic age is the year gap between the year of an author's first paper recorded in the PKG dataset and the publication year of the focal paper. We first calculated the academic age of each thought leader in a team and then computed the average academic age across all team thought leaders. Previous studies have used academic age to proxy research experience (AlShebli et al. 2018) and leadership power (Xu, Bu, et al. 2022).

**Average prior productivity of thought leader(s).** An author's prior productivity is counted as the accumulated number of publications that preceded the year in which the paper was published. Similarly, we first counted the prior productivity of each thought leader in a team and then calculated the average prior productivity across all team thought leaders. Average prior productivity is a direct indicator of research ability, which tends to affect team performance (Huo et al. 2019).

### 3.2.3 Multivariable regression

Multivariable regression was used to quantify the relationship between the number of thought leaders and team performance, i.e., addressing *RQ1*. The empirical model is defined as follows:

$$teamPerformance_{ij} = \beta_0 + \beta_1 RTL_{ij} + \beta_2\left(RTL_{ij}^2\right) + \beta_3 X_{ij} + D_{ij} + Y_{ij} + \varepsilon, i \in \{0,1\} \quad (2)$$

Where $teamPerformance_{ij}$ represents the team impact or team disruptiveness of team $j$, $i = 0$ for team impact, and $i = 1$ for team disruptiveness. We employed the natural logarithm transformation of citation counts for regression, as recommended by Thelwall and Wilson (2014). Considering the presence of zero-citations articles, we also added one to the original citation counts before the natural logarithm transformation. $RTL_{ij}$ is the ratio of thought leaders of team *j*. To model a possible non-linear relationship between the ratio of thought leaders and team performance, the square term of the ratio of thought leaders (i.e., $RTL_{ij}^2$) was included in our model. $X_{ij}$ is a set of control variables of team $j$. $D_{ij}$ and $Y_{ij}$ represent the publication year and disciplines of article team $j$ respectively, and we included these two types of dummy variables to control for the fixed effect of time and discipline. Finally, ordinary least squares (OLS) was adopted to estimate the coefficients of the model.

#### 3.2.4 Coarsened exact matching

Coarsened exact matching (CEM) was adopted to gain further insight into the causal relationship between the number of thought leaders and team performance (answering *RQ2*). CEM has been widely applied to observational data to infer causality (Iacus et al. 2012). Specifically, CEM matches the control and treatment groups in terms of the covariates identified, thus reducing the confounding effect. In practice, CEM coarsens each of $n$ covariates into $m_i$ bins, producing $\prod_{i=1}^{n} m_i$ strata. Then, each unit is assigned a stratum based on the corresponding value of covariates. Only units in any stratum that contain at least one control and treated unit are retained. We adopted k-to-k solution to avoid using weights in the inference of average treatment effect (ATT), as recommended by Ho et al. (2007). In brief, we applied the nearest neighbor selection method within each stratum to ensure an equal number of treated and control units in all strata, thereby achieving balanced matching results (Ho et al. 2007).

In our context, when studying the effect of number of thought leaders on team performance, we used the following strategy to generate treatment and control groups:

$$\begin{cases} treatment\ group: RTL > P_{1-i} \\ control\ group: RTL \le P_i \end{cases} \quad (3)$$

Where $P_i$ represents thresholds of $RTL$ (Ratio of thought leaders), with $i \in \{0.1, 0.2, 0.3, 0.4, 0.5\}$ and used to divide groups. CEM was run five times (one for each threshold). The treatment group denotes the teams with larger RTL, and if the number of thought leaders is associated with team performance, we expect a significant difference in team performance between the two groups.

## 4. Results

To address *RQ1*, we first investigated the relationship between the number of thought leaders and team performance from an exploratory point of view (Section 4.1). Next, we utilized multivariable regression to further quantify this correlational relationship (Section 4.2). Then, we examined the potential causal effect of the number of thought leaders on team performance to answer *RQ2* (Section 4.3). Finally, we conducted the robustness checks (Section 4.4).

### 4.1 Descriptive statistics

Figure 2 depicts the number of thought leaders against team performance. The *horizontal* axis denotes the number of thought leaders, the left *vertical* axis (in light red) represents team impact (i.e., citation counts), and the right *vertical* axis (in light blue) represents team disruptiveness (i.e., disruption percentile). The colored numbers above the curves are the slope of fitted linear regression lines. We only visualized the teams with a size smaller than or equal to 10, which account for approximately 85% of all teams. The distribution of the number of thought leaders across team sizes is presented in Figure S1. As a general trend, the larger the number of thought leaders, the more impactful and less disruptive the scientific outputs produced by teams (See Figure 2). More specifically, the fitted regression slope between the number of thought leaders and team impact is significantly positive in some partial cases, while the relationship between the number of thought leaders and disruptiveness shows a significantly negative slope across different team sizes. The fitted slope of the number of thought leaders and team impact is insignificantly negative when the team size is 3, 4, and 6, but the curve shows an inverted U-shape. For the teams with a size of 3 (see Figure 2B), team impact first goes up and then goes down as the number of thought leaders increases, and the maximum team impact is achieved when the number of thought leaders is 2. Noteworthy, we only consider a linear function to fit the data, which may be one reason for the insignificantly negative relationship.

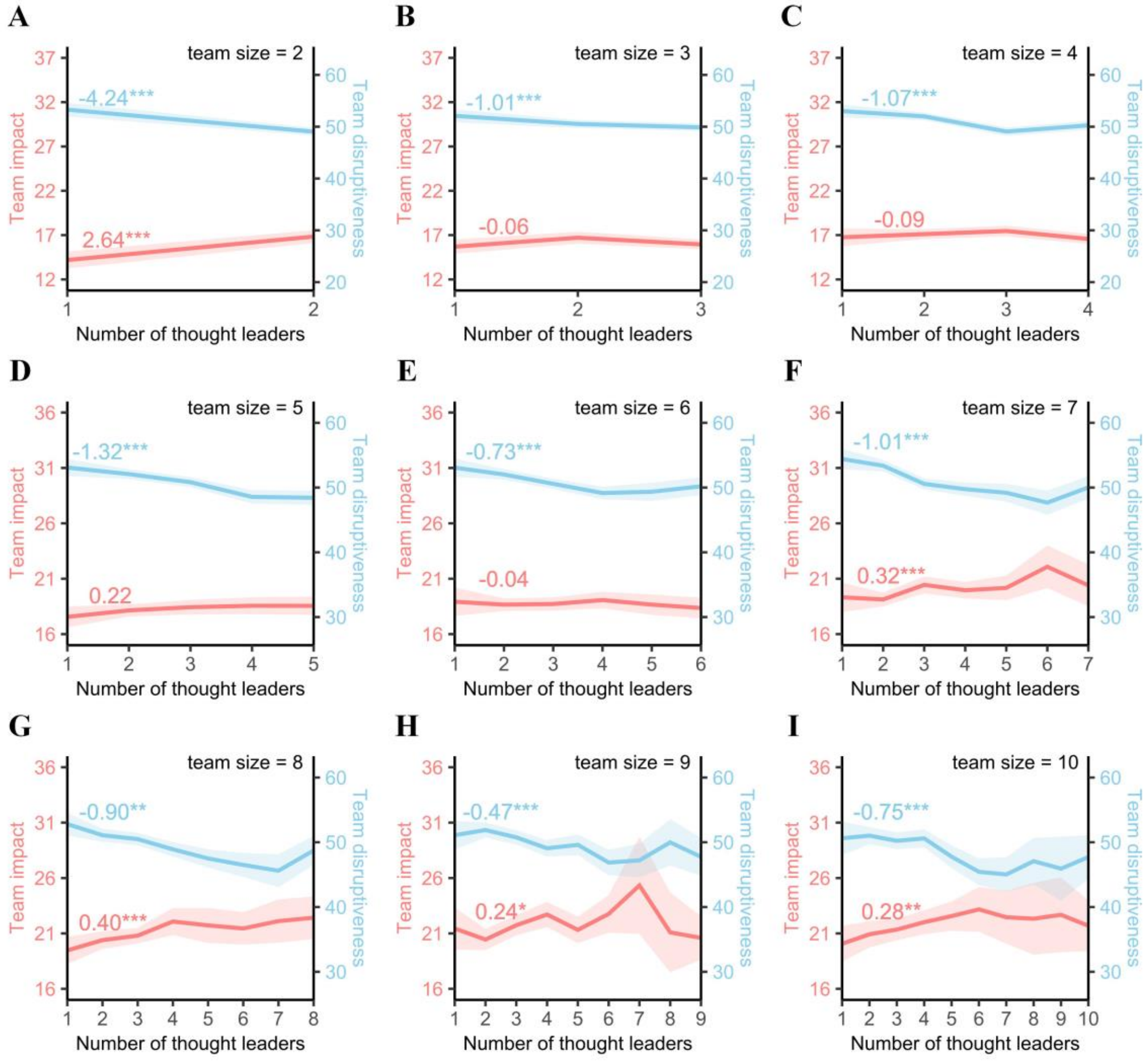

Figure 2 Number of thought leaders and team performance by team sizes ($^{***}$ $p < 0.01$, $^{**}$ $p < 0.05$, $^{*}$ $p < 0.1$.)

In addition, we further investigated the relationship between the number of thought leaders and team performance at an aggregate level (Figure 3). More exactly, teams are binned into five bins (i.e., 0-0.2, 0.2-0.4, 0.4-0.6, 0.6-0.8, 0.8-1.0) based on the value of their ratio of thought leaders. As shown in Figure 4, team impact and team disruptiveness decreased with an increase in the ratio of thought leaders. The relationship between team disruptiveness and the ratio of thought leaders aligns with the observations presented in Figure 2. However, the association between team impact and the ratio of thought leaders remained ambiguous, suggesting the need for further multivariate analysis and causal analysis.

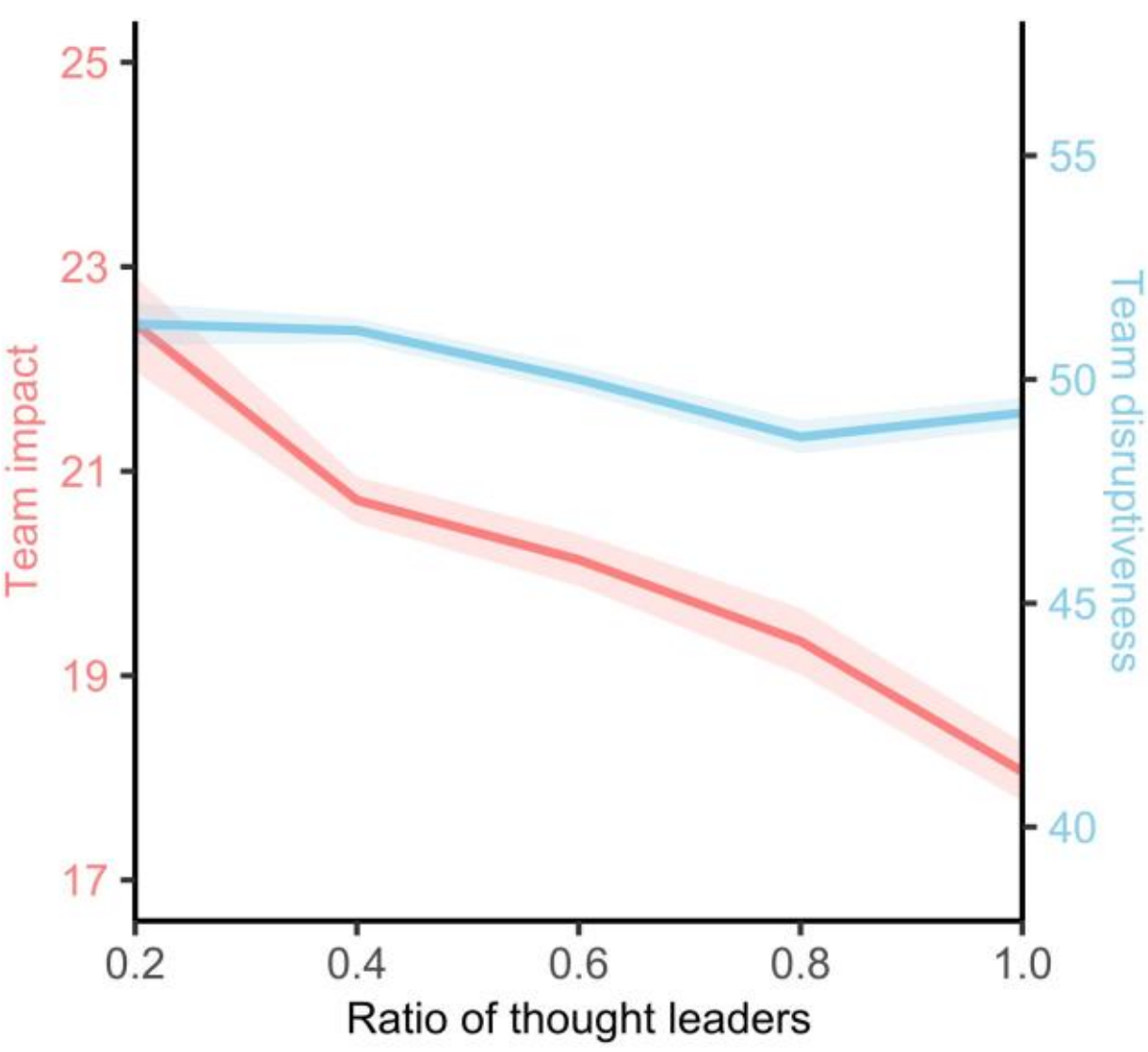

Figure 3 The relationship between the ratio of thought leaders and team performance

The descriptive statistics for all variables and correlations between them are reported in Table 2. The descriptive statistics of *team impact* and *team disruptiveness* indicate that, on average, the team impact and disruptiveness were 19.99 and 50.08, respectively. The mean value of the *ratio of thought leaders* was 0.54, suggesting that scholars commonly serve as thought leaders in scientific teams. Team size ranged from 2 to 200, with a mean of 7.08, indicating that large teams tended to dominate the knowledge production in our sample. In addition, 35% of the teams were international teams, and 86% were mixed-gender teams. Most correlations are considered significant at a p-value < 0.05. As expected, *team impact* was negatively correlated with *team disruptiveness* ($r$ = -0.30). In addition, the *ratio of thought leaders* was negatively correlated with both *team impact* ($r$ = -0.05) and *team disruptiveness* ($r$ = -0.03), which is consistent with the observations depicted in Figure 3. Moreover, the correlations among these independent variables did not indicate potential multicollinearity. For more robust results, we conducted a multicollinearity test before the multivariate regression. The variance inflation factor (VIF) of all independent variables was less than 10, and the mean VIF was 4.05, indicating no severe multicollinearity.

Table 2 Descriptive statistics and correlations

| No. | Variable | Mean | S.D. | Min | Median | Max | Correlation | | | | | | | | | | |
|---|---|---|---|---|---|---|---|---|---|---|---|---|---|---|---|---|---|
| | | | | | | | 1 | 2 | 3 | 4 | 5 | 6 | 7 | 8 | 9 | 10 | 11 |
| 1 | Team impact | 19.99 | 24.80 | 0 | 14 | 1186 | 1 | | | | | | | | | | |
| 2 | Team disruptiveness | 50.08 | 28.69 | 0.10 | 50.2 | 100 | **-0.30** | 1 | | | | | | | | | |
| 3 | Ratio of thought leader(s) | 0.54 | 0.27 | 0.02 | 0.5 | 1 | **-0.05** | **-0.03** | 1 | | | | | | | | |
| 4 | Team size | 7.08 | 4.23 | 2 | 6 | 200 | **0.16** | **-0.04** | **-0.44** | 1 | | | | | | | |
| 5 | Institutional diversity | 2.69 | 1.92 | 1 | 2 | 62 | **0.11** | **-0.05** | **-0.21** | **0.63** | 1 | | | | | | |
| 6 | International team | 0.35 | 0.48 | 0 | 0 | 1 | **0.05** | **-0.07** | **-0.09** | **0.23** | **0.44** | 1 | | | | | |
| 7 | Number of references | 42.07 | 20.22 | 5 | 40 | 798 | **0.18** | **0.01** | **-0.05** | **0.08** | **-0.01** | **-0.01** | 1 | | | | |
| 8 | Gender diversity | 0.86 | 0.34 | 0 | 1 | 1 | **0.04** | 0.00 | **-0.24** | **0.28** | **0.15** | **0.07** | **0.06** | 1 | | | |
| 9 | Publication year | 2012.72 | 1.90 | 2007 | 2013 | 2015 | **-0.18** | **0.18** | **-0.01** | **0.02** | 0.000 | **-0.02** | **-0.06** | **0.04** | 1 | | |
| 10 | Avg. academic age of thought leader(s) | 14.04 | 7.33 | 0 | 13 | 69 | **0.04** | **-0.06** | **-0.10** | **0.13** | **0.09** | **0.03** | **0.13** | **0.05** | 0.00 | 1 | |
| 11 | Avg. prior productivity of thought leader(s) | 52.93 | 60.11 | 0 | 35.75 | 1333 | **0.06** | **-0.01** | **-0.13** | **0.18** | **0.12** | **0.04** | **0.03** | **0.08** | **0.055** | **0.51** | 1 |

**Notes**: correlations with bold numbers at $p<0.05$.

### 4.2 The regression results

To answer *RQ1*, we performed a series of multivariable regressions to examine the effect of the number of thought leaders on two dimensions of team performance: team impact and team disruptiveness. The regression results are presented in Table 3. We included the independent variable only in Models (1) and (5); the estimates indicated that the ratio of thought leaders is negatively associated with both team impact ($\beta = -0.269, p < 0.01$) and team disruptiveness ($\beta = -2.849, p < 0.01$), resonating with the results depicted in Figure 4. Models (2) and (6) served as the baseline models, and we included only control variables in these two models. When adding control variables in Models (3) and (7), the effect of the ratio of thought leaders on team impact became significantly positive ($\beta = 0.030, p < 0.01$) compared to that in Model (1). In contrast to Model (5), its influence on team disruptiveness remained significantly negative ($\beta = -5.570, p < 0.01$). We also noticed that the Adj R-squared in Model (3) (0.1531) was larger than that in Model (1) (0.0074), suggesting that adding control variables to the model would improve the goodness of fit. For team disruptiveness, the results were the same (see Adj R-squared in Models (5) and (7)).

As observed and discussed earlier, we also examined the presence of curvilinear relationships between the ratio of thought leaders and team performance. In Models (4) and (8), the ratio of thought leaders and its square term were introduced. The coefficients for the ratio of thought leaders ($\beta = 0.213, p < 0.01$) and its square term ($\beta = -0.152, p < 0.01$) were both significant in Model (4), implying an inverted-U-shaped association between the ratio of thought leaders and team impact. We further found that team impact was maximized when the ratio of thought leaders was 0.70, as shown in Figure 4A, indicating that the net effect of the number of thought leaders on team impact becomes significantly negative when teams have too many thought leaders. When the data beyond the estimated turning point (maximum of the curve) are limited, the quadratic function often implies inverse-U-shaped relationships that do not truly exist (Simonsohn 2019). Our study included 39,040 teams (27.58% of the total) beyond the turning point, reassuring the robustness of the curvilinear association. Similarly, for team disruptiveness, the estimates for the ratio of thought leaders ($\beta = -12.700, p < 0.01$) and its square term ($\beta = 5.932, p < 0.01$) were both significant in Model (8). However, the estimated turning point ($\hat{x} = 1.070$) was beyond the upper bound of the ratio of thought leaders, suggesting that U-shaped relationships do not exist and the ratio of thought leaders had a negative association with the disruptive percentile (See Figure 4B).

The regression coefficients for the control variables also yielded interesting results. First, team size was positively linked to team impact ($\beta = 0.032, p < 0.01$, in Model (3)) and negatively linked to disruptiveness ($\beta = -0.362, p < 0.01$, in Model (7)). These results align with previous findings that small teams tended to disrupt prior knowledge and technology, whereas large teams were more likely to produce incremental ideas (Wu et al. 2019). Second, aligning with a prior study (Wagner et al. 2019), we found that international teams were less likely to propose disruptive ideas ($\beta = -3.299, p < 0.01$, in Model (7)), but may receive more citations ($\beta = 0.038, p < 0.01$, in Model (3)). Third, the number of references was positively associated with team impact ($\beta = 0.010, p < 0.01$, in Model (3)) and disruptiveness ($\beta = 0.032, p < 0.01$, in Model (7)), confirming previous findings in the literature (Ruan et al. 2021; Wei et al. 2023). Fourth, gender diversity was positively correlated with team impact ($\beta = 0.037, p < 0.01$, in Model (3)) but negatively associated with team disruptiveness ($\beta = -0.005$, in Model (7)), albeit non-significantly. Finally, for thought leaders' characteristics, academic age negatively affected team impact ($\beta = -0.003, p < 0.01$, in Model (3)) and disruptiveness ($\beta = -0.272, p < 0.01$, in Model (7)), while prior productivity positively affected team impact ($\beta = 0.001, p < 0.01$, in Model (3)) and team disruptiveness ($\beta = 0.011, p < 0.01$, in Model (7)). Although academic age and productivity are often used as proxies for an individual's status or leadership power in teams (Xu, Bu, et al. 2022), their essences diverge. According to Prato et al. (2019), academic age represents a type of ascribed status, whereas productivity is a type of achieved status, and the effects of each on team performance may differ.

Table 3 The association between the number of thought leaders and team performance

| Variables \ Model No. | (1) Team impact | (2) Team impact | (3) Team impact | (4) Team impact | (5) Team disruptiveness | (6) Team disruptiveness | (7) Team disruptiveness | (8) Team disruptiveness |
|---|---|---|---|---|---|---|---|---|
| RTL | -0.269*** | | 0.030*** | 0.213*** | -2.849*** | | -5.570*** | -12.700*** |
| | (0.008) | | (0.010) | (0.039) | (0.279) | | (0.313) | (1.336) |
| RTL squared | | | | -0.152*** | | | | 5.932*** |
| | | | | (0.031) | | | | (1.085) |
| Team size | | 0.031*** | 0.032*** | 0.032*** | | -0.201*** | -0.362*** | -0.378*** |
| | | (0.001) | (0.001) | (0.001) | | (0.024) | (0.026) | (0.026) |
| Institutional diversity | | -0.001 | -0.002 | -0.002 | | -0.024 | 0.060 | 0.059 |
| | | (0.002) | (0.002) | (0.002) | | (0.056) | (0.057) | (0.057) |
| International team | | 0.037*** | 0.038*** | 0.037*** | | -3.222*** | -3.299*** | -3.256*** |
| | | (0.005) | (0.005) | (0.005) | | (0.177) | (0.177) | (0.177) |
| Number of references | | 0.010*** | 0.010*** | 0.010*** | | 0.032*** | 0.032*** | 0.032*** |
| | | (0.000) | (0.000) | (0.000) | | (0.004) | (0.004) | (0.004) |
| Gender diversity | | 0.035*** | 0.037*** | 0.035*** | | 0.492** | -0.005 | 0.093 |
| | | (0.007) | (0.007) | (0.007) | | (0.234) | (0.236) | (0.237) |
| Avg. academic age of TLs | | -0.003*** | -0.003*** | -0.003*** | | -0.269*** | -0.272*** | -0.271*** |
| | | (0.000) | (0.000) | (0.000) | | (0.012) | (0.012) | (0.012) |
| Avg. prior productivity of TLs | | 0.001*** | 0.001*** | 0.001*** | | 0.011*** | 0.011*** | 0.011*** |
| | | (0.000) | (0.000) | (0.000) | | (0.001) | (0.001) | (0.001) |
| Constant | 2.817*** | 2.534*** | 2.510*** | 2.468*** | 51.631*** | 38.757*** | 43.144*** | 44.798*** |
| | (0.005) | (0.022) | (0.024) | (0.025) | (0.170) | (0.707) | (0.749) | (0.806) |
| Field-fixed effect | No | Yes | Yes | Yes | No | Yes | Yes | Yes |
| Year-fixed effect | No | Yes | Yes | Yes | No | Yes | Yes | Yes |
| Observations | 141,550 | 141,550 | 141,550 | 141,550 | 141,550 | 141,550 | 141,550 | 141,550 |
| Adj R-squared | 0.0074 | 0.1530 | 0.1531 | 0.1532 | 0.0007 | 0.0455 | 0.0477 | 0.0479 |

**Notes**: TL denotes thought leader; RTL denotes the ratio of thought leaders; Standard errors in parentheses. *** $p < 0.01$, ** $p < 0.05$, * $p < 0.1$.

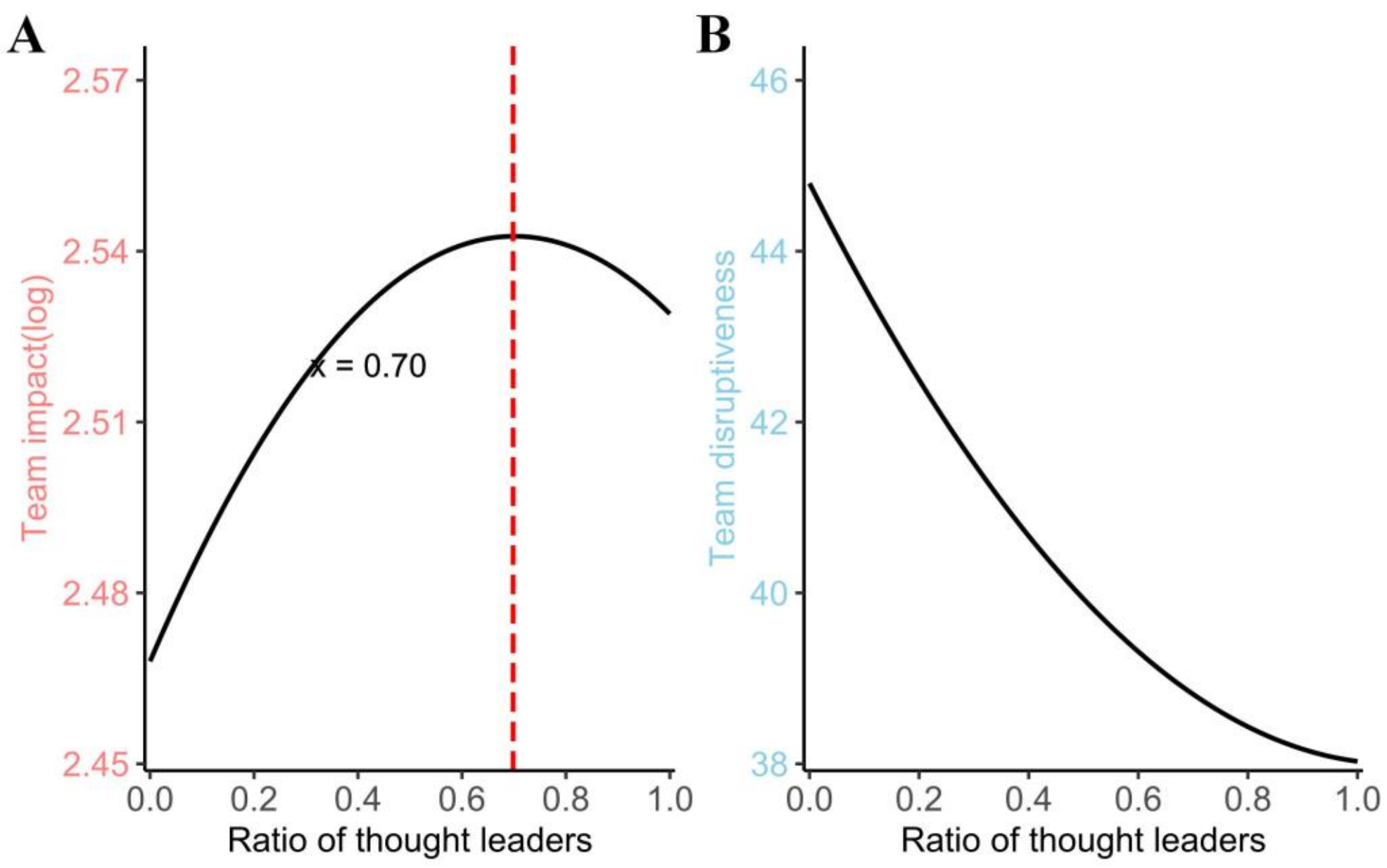

Figure 4 The estimated relationship between the ratio of leaders and team performance

### 4.3 The potential causal effect of the number of thought leaders on team performance

Although several variables were included to control for the confounding factors in the regression model, we cannot interpret our findings as causal. Because the regression model was based on an unbalanced dataset and non-overlapping distribution of the control variables, we could not eliminate the effects of all possible confounding factors, and the results obtained from the previous regression analysis may suffer from estimation bias. For example, when the team size was 2, there were 1,905 teams with one thought leader and 5,651 teams with two thought leaders (see Figure s1A). To mitigate the potential bias and infer causality, we adopted CEM to further investigate the relationship between the number of thought leaders and team performance, answering *RQ2*. CEM simulates a quasi-experimental design to infer causality (Iacus et al. 2012) and has been widely used in the science of science field (AlShebli et al. 2018; Shi et al. 2023).

The essential reason for applying CEM was to find team pairs that differed in the treatment (i.e., number of thought leaders) but were similar or the same in thought leader group and team level characteristics. In our study, the treatment and control groups refer to the teams with a ratio of thought leaders higher or lower than a certain threshold (See section 3.2.4 for details). The covariates used in CEM were team size, international team, number of references, gender diversity, publication year, research field, average academic age of thought leaders, and average prior productivity of thought leaders. Institutional diversity was not included in the covariates set because it is irrelevant to team performance according to our regression results (see Table 3). Before performing the matching, CEM was required to coarsen each variable into different bins. The binning criteria of each variable are shown in Table 4.

Table 4 The binning strategy of covariates

| Variables | # of bins | Binning strategy |
|---|---|---|
| **Team Level** | | |
| Team size | 91 | One for each team size |
| International team | 2 | 1: international team; 0: domestic team |
| Number of references | 4 | Quantile-based binning: 5-28, 29-40, 41-53, >53 |
| Gender diversity | 2 | 1: mixed-gender team; 0: same-gender team |
| Publication year | 9 | Each bin corresponds to a single year between 2007 and 2015 |
| Research field | 14 | Each bin represents one discipline |
| **TL group** | | |
| Avg. academic age of TL(s) | 6 | Equal-interval binning: 0-11.5, 11.6-23.0, 23.1-34.5, 34.6-46.0, 46.1-57.5, >57.5 |
| Avg. prior productivity of TL(s) | 4 | Quantile-based binning: 0-16.75, 16.76.-35.75, 35.76-68, >68 |

The exact matching method was employed for the coarsened data to select the matched units. The covariates balance check for CEM5 (T: RTL >0.5; C: RTL <=0.5) is shown in Table 5. The standardized bias of all covariates was significantly reduced after matching. Before matching, all covariates exhibited significant differences in mean values between the treatment and control groups. After matching, this difference disappeared, which implies that the treatment and control groups are comparable in all covariates. Similarly, other CEM groups also showed good matching results. The results of the covariates balance check are shown in Table s1–s4.

Drawing on the CEM results, the ATT for the five groups was estimated. The results are presented in Table 6. In general, as the gap in the ratio of thought leaders between treatment and control groups increased, the difference in team impact and disruptiveness between the two groups increased. Nevertheless, the difference did not increase after reaching a peak value. Specifically, in the case of teams with a ratio of thought leaders above 0.8, the ATT in scientific impact reached the peak of 0.858 ($p < 0.1$, in CEM (2)), compared to their counterparts, with a ratio of thought leaders equal to or below 0.2. However, when comparing the teams with RTL larger than 0.9 versus the teams with RTL smaller than or equal to 0.1, the estimated ATT becomes negative, albeit non-significantly ($ATT = -0.641, p > 0.1$, in CEM (1)).

The results partly demonstrate that the number of thought leaders has a non-linear relationship with team impact, which means that increasing the number of thought leaders in a team does not necessarily improve team impact. Interestingly, a similar pattern was observed for team disruptiveness. Aside from CEM1, all ATTs for team disruptiveness were significant. The ATT for the team disruptiveness in CEM2 was -5.082, which means that teams with RTL above 0.8 could reduce 5.082 disruption percentiles relative to those teams with RTL equal to or below 0.2. When the RTL between treatment and control groups was further increased, the ATT remained negative but insignificant. Teams with fewer thought leaders were more prone to producing disruptive outputs than those with more thought leaders.

Table 5 The covariates balance check for the CEM5 model (T: RTL >0.5; C: RTL <=0.5)

| Variable | Before matching | | | | After matching | | | | %bias (reduction) |
|---|---|---|---|---|---|---|---|---|---|
| | T(Mean) | C(Mean) | % bias | p-value | T(Mean) | C(Mean) | % bias | p-value | |
| Team size | 5.562 | 8.271 | -32.754 | 0 | 6.531 | 6.531 | 0.000 | 1.000 | 100 |
| International team | 0.321 | 0.369 | -12.974 | 0 | 0.333 | 0.333 | 0.000 | 1.000 | 100 |
| Number of references | 41.244 | 42.709 | -3.429 | 0 | 42.130 | 42.076 | 0.128 | 0.717 | 96.267 |
| Gender diversity | 0.799 | 0.916 | -12.731 | 0 | 0.901 | 0.901 | 0.000 | 1.000 | 100 |
| Publication year | 2012.691 | 2012.750 | -0.003 | 0 | 2012.831 | 2012.831 | 0.000 | 1.000 | 100 |
| Avg. academic age of TL(s) | 13.443 | 14.501 | -7.296 | 0 | 13.538 | 13.486 | 0.391 | 0.282 | 94.641 |
| Avg. prior productivity of TL(s) | 46.004 | 58.366 | -21.180 | 0 | 47.658 | 47.737 | -0.167 | 0.825 | 99.212 |

**Notes**: TL denotes thought leader.

Table 6 The causal effects of TL size on team performance

| Model | Groups | $\lvert T^{before} \rvert$ | $\lvert C^{before} \rvert$ | $\lvert T^{after} \rvert$ | $\lvert C^{after} \rvert$ | ATT (team impact) | $CI_{0.95}$ | P-value | ATT (team disruptiveness) | $CI_{0.95}$ | P-value |
|---|---|---|---|---|---|---|---|---|---|---|---|
| CEM1 | T: RTL >0.9<br>C: RTL <=0.1 | 24008 | 1817 | 234 | 234 | -0.641 | [-4.455,3.173] | 0.742 | -4.097 | [-9.098,0.903] | 0.110 |
| CEM2 | T: RTL>0.8<br>C: RTL <=0.2 | 27659 | 14066 | 4186 | 4186 | 0.858 | [-0.122,1.838] | 0.086 | -5.082 | [-6.306,-3.858] | 0 |
| CEM3 | T: RTL >0.7<br>C: RTL <=0.3 | 39040 | 31558 | 10462 | 10462 | 0.755 | [0.140,1.370] | 0.016 | -3.915 | [-4.691,-3.139] | 0 |
| CEM4 | T: RTL >0.6<br>C: RTL <=0.4 | 51037 | 54811 | 21287 | 21287 | 0.754 | [0.316,1.193] | 0 | -3.163 | [-3.705,-2.622] | 0 |
| CEM5 | T: RTL >0.5<br>C: RTL <=0.5 | 62211 | 79339 | 35063 | 35063 | 0.490 | [0.154,0.825] | 0.004 | -2.352 | [-2.774,-1.930] | 0 |

**Notes**: | | represents the sample size; RTL denotes the ratio of thought leaders.

### 4.4 Robustness checks

Table 1 shows that approximately 10% of the publications were not assigned to any discipline of WoS. In the analysis, we categorized these publications into one discipline, namely “other discipline.” This discipline may include publications from multiple disciplines. Though including these publications in the analysis resulted in a larger dataset, this data processing strategy may yield biased findings. We excluded publications without identifiable disciplines and repeated the empirical analysis to ensure the robustness of our findings. The results of multivariate linear regression shown in Table s5 indicate that the number of thought leaders has an inverted-U-shaped relationship with team impact and is negatively associated with team disruptiveness. The findings are consistent with previous results in Table 3. As displayed in Table s6, the causal inference results also align with the conclusion presented in Table 6. Moreover, to mitigate potential bias in assessing the impact of article teams, we adopted a five-year citation count that is normalized for publication year and discipline for a robustness check, following the practice of Yang, Tian, et al. (2022). We conducted a similar regression by replacing the five-year citation count variable with this normalized one, and the main finding displayed in Table s7 is consistent with the results reported in Table 3.

## 5. Discussion

Thought leadership is important in facilitating innovation (McCrimmon 2005). This study investigated the effect of the number of thought leaders on two dimensions of team performance from correlational and causal perspectives.

The result showed an inverted-U-shaped relationship between the number of thought leaders and team impact. This finding is consistent with the results from Swaab et al. (2014) and Szatmari (2022), though these studies were conducted on sports teams. One explanation for this non-linear relationship is that the audience effect and team conflicts may simultaneously affect team impact. At the initial stage, as the number of thought leaders increases, the audience effect plays a major role in attracting attention to research (Thelwall et al. 2023). During this stage, only a small proportion of team members act as thought leaders, and those thought leaders have strong belongingness and satisfaction and are thus fully engaged in boosting innovation (Hogg 2001). After the ratio of thought leaders exceeds a certain point, most of the team members serve as thought leaders in teams, and belongingness and satisfaction from playing a thought leader role in teams decrease. Those thought leaders may turn to acquire status, while team leaders will put forth greater efforts to consolidate their position in the status hierarchy (Pettit et al. 2010), which may result in team status conflicts. Consequently, team status conflicts hinder decision-making effectiveness and produce low-quality ideas. The positive effect of the audience effect may be offset by the negative effect induced by team status conflicts (Lovelace et al. 2001), resulting in a net negative effect.

The results also showed that teams with more thought leaders tended to produce less disruptive ideas. This finding corroborates the findings of Wu et al. (2019), even though their study focuses on team size without distinguishing between roles in teams. Moreover, our findings align with previous experimental research that found individuals tend to produce fewer ideas than teams (Paulus et al. 2013) and neutralize one another's viewpoints (Greenstein & Zhu 2016).

Additional analyses also led to interesting conclusions. First, we observed that international collaboration improved team impact but hindered team disruptiveness, perhaps because international collaborative publications are likely to expand the scope of international audiences and, therefore, receive more citations than studies without international collaboration. This finding aligns with previous research (Fan et al. 2022). Lin et al. (2023) suggested that remote collaboration is less likely than on-site collaboration to produce breakthrough discoveries, which aligns with our results.

### 5.1 Theoretical and practical implications

The theoretical contributions of this study are threefold. First, this study extends the concept of thought leadership from organizational management into the science of team science domain. Scientific teams operate on the impetus of innovation and knowledge, which might be distinguished from profit-driven firms and marketing (Harvey et al. 2021; Neuhaus et al. 2022). Second, this study provides new empirical evidence to understand the number of thought leaders within scientific teams and its impact on two aspects of team performance: team impact and disruptiveness. This evidence enriches the literature on team structure and team performance. Third, this study pioneers the application of CEM to investigate the potential causal effect of the number of thought leaders on team performance.

Our findings also have practical implications for policymakers and team managers. From a public policy standpoint, funding aimed at disruptive innovation is best assigned to teams with a small number of thought leaders, while funding aimed at high-impact and incremental innovation should be allocated to teams with a large number of thought leaders. For team managers seeking to boost their teams' impact, keeping the ratio of thought leaders within an optimal range is crucial, as having too many or too few thought leaders is not beneficial. If the team aims to achieve breakthrough innovation, a small number of thought leaders is preferable.

### 5.2 Limitations

This study had several limitations. First, the dataset mainly contained publications in the natural sciences. Given that research paradigms differ across disciplines, these findings may not generalize to other disciplines, especially social sciences, arts, and humanities. Second, although CEM simulates a quasi-experimental design to infer causality, it is not a perfect substitute for randomized controlled experiments; unobserved confounding factors may remain. Thus, future studies could include more available confounding factors or use other causal inference methods, such as instrumental variables estimations. Third, we only consider two widely used aspects to evaluate team performance: team impact and disruptiveness. Nevertheless, it would be interesting to measure team performance from other aspects, such as multidisciplinary impact (Bu et al. 2021), interdisciplinarity (Leahey et al. 2016), and scientific novelty (Uzzi et al. 2013). Fourth, we defined the authors responsible for conceptual tasks as thought leaders and identified those involved in the task of “conceived and designed the experiments” as thought leaders. However, we must acknowledge that thought leaders in scientific teams could be defined and operationalized in different ways. Fifth, we need to acknowledge that a very small fraction of papers may disclose the author contribution statements in an arbitrary manner, potentially biasing the identification of thought leaders within teams. Sixth, we include several important confounding factors that are easy to capture and control in the regression models. However, there may be other influential factors that are omitted, potentially biasing the results.

## 6. Conclusions

This study aimed to examine the relationship between the number of thought leaders and two aspects of team performance: team impact and disruptiveness. Based on a dataset comprising 141,550 articles with parsed author contribution statements, we defined the authors who contributed to the conception of ideas in teams as *thought leaders*. We then explored patterns related to the number of thought leaders and team performance across team sizes and investigated the effects of the number of thought leaders on team performance.

First, the descriptive results showed that teams with more thought leaders tended to produce more impactful and less disruptive scientific outputs. Second, in the empirical model, we controlled various confounding factors and observed an inverted-U-shaped relationship between the number of thought leaders and team impact. Specifically, at the initial stage, a high ratio of thought leaders positively affects team impact. Nevertheless, once the ratio of thought leaders reaches 70%, a higher proportion of thought leaders negatively affects team impact. For disruptiveness, we found that the number of thought leaders is negatively associated with the disruptiveness of team outputs.

Furthermore, we examined the effect of the number of thought leaders on team performance from a causal perspective by applying CEM. The results suggested that the number of thought leaders increases team impact but decreases team disruptiveness. In addition, as the RTL gap between the treatment and control groups increased, the difference in scientific impact and disruptiveness between the two groups broadened. However, the difference did not increase after RTL reached a peak value. In other words, the effect of the number of thought leaders on team performance was non-linear. In summary, more heads do not always imply better performance.

This study provides several promising directions for future research. Firstly, future studies should expand the empirical study to include a larger sample size. For example, prestigious journals, including Nature, Science, and PNAS, also require the authors to disclose individual authors’ contributions (Sauermann & Haeussler 2017). While these journals do not provide author contributions statements in a machine-readable format, web crawler technology could be employed to extract the data. Secondly, to ensure a more robust result, additional individual-level confounding factors could be included in empirical models, such as thought leaders’ personalities, leadership styles, and so on. Thirdly, while we only demonstrated that a large number of thought leaders leads to a decrease in the disruptiveness of team outputs, the possible mechanisms and potential mitigating factors behind this relationship deserve deep exploration. Fourthly, it would be interesting to investigate the representation of women thought leaders in scientific teams and explore the association between the gender diversity in thought leadership and team performance.

**Acknowledgements**

This work is supported by the National Research Foundation of Korea (NRF) grant funded by the Korea government (MSIT) (No. RS-2023-00209775), Jiangsu Social Science Fund (Grant No. 20TQA001), National Natural Science Foundation of China (Grant No. 72004054 and 72074113) and China Scholarship Council (ID: 202206840067). The authors would like to thank NamSor company for providing the free services of name-to-gender inference. We also thank the two anonymous reviewers for improving the quality of this article. Yi Zhao would also like to give special thanks to his wife, Yi Zhao (Mia), for her support and understanding during his one-year visit to Yonsei University.

## Appendix

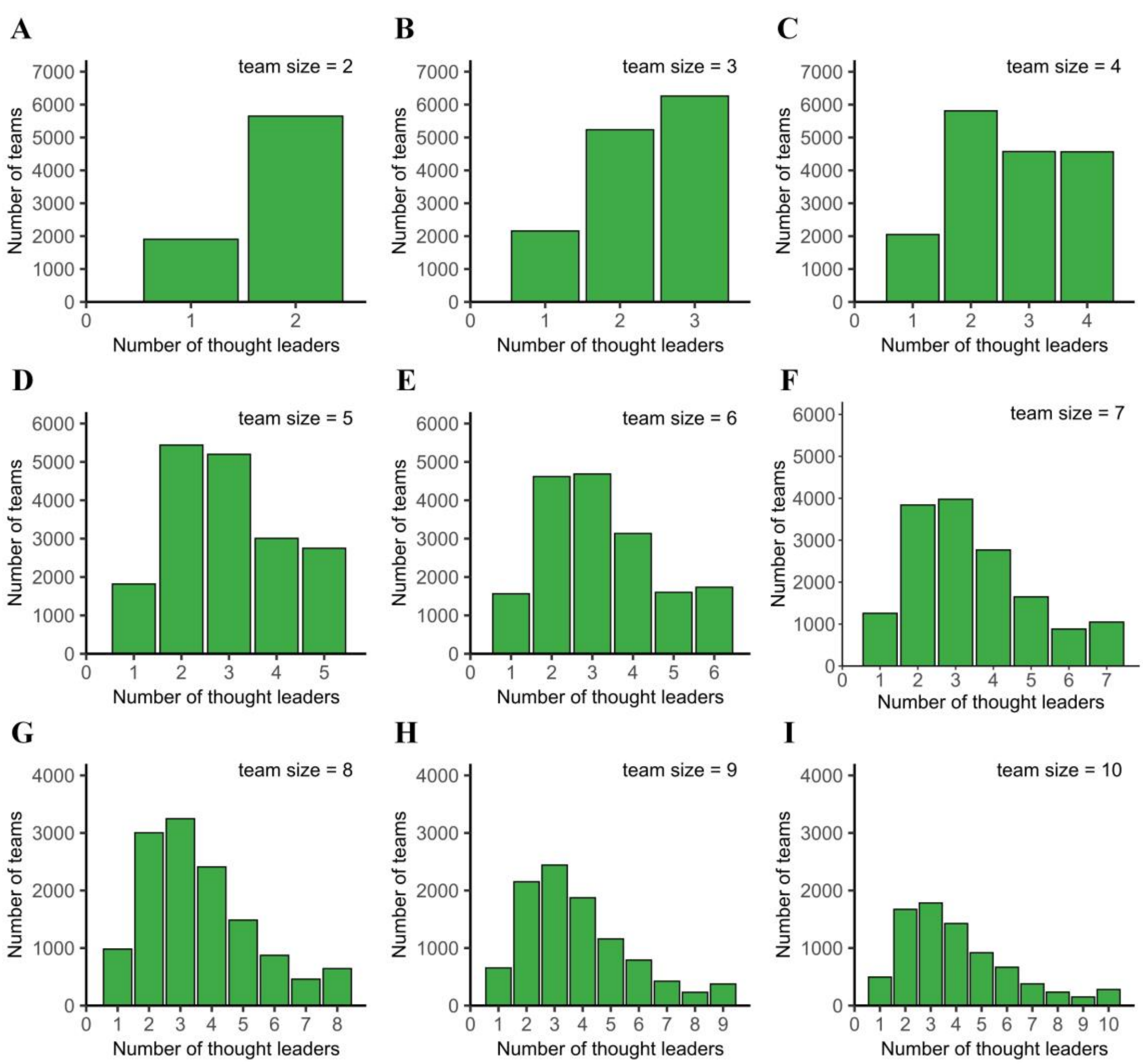

Figure s1 The distribution of the number of thought leaders across team sizes

Table s1 The covariates balance check for the CEM1 model (T: RTL >0.9; C: RTL <=0.1)

| Variable | Before matching | | | | After matching | | | | %bias (reduction) |
|---|---|---|---|---|---|---|---|---|---|
| | T(Mean) | C(Mean) | % bias | p-value | T(Mean) | C(Mean) | % bias | p-value | |
| Team size | 4.199 | 15.554 | -73.003 | 0.000 | 11.231 | 11.231 | 0.000 | 1.000 | 100 |
| International team | 0.250 | 0.489 | -48.852 | 0.000 | 0.444 | 0.444 | 0.000 | 1.000 | 100 |
| Number of references | 39.720 | 42.964 | -7.549 | 0.000 | 43.509 | 42.744 | 1.790 | 0.632 | 76.288 |
| Gender diversity | 0.688 | 0.987 | -30.239 | 0.000 | 0.991 | 0.991 | 0.000 | 1.000 | 100 |
| Publication year | 2012.725 | 2012.637 | 0.004 | 0.068 | 2013.295 | 2013.295 | 0.000 | 1.000 | 100 |
| Avg. academic age of TL(s) | 12.577 | 15.757 | -20.180 | 0.000 | 13.688 | 13.444 | 1.808 | 0.601 | 91.041 |
| Avg. prior productivity of TL(s) | 39.610 | 70.412 | -43.746 | 0.000 | 53.253 | 57.034 | -6.629 | 0.401 | 84.847 |

Table s2 The covariates balance check for the CEM2 model (T: RTL >0.8; C: RTL <=0.2)

| Variable | Before matching | | | | After matching | | | | %bias (reduction) |
|---|---|---|---|---|---|---|---|---|---|
| | T(Mean) | C(Mean) | % bias | p-value | T(Mean) | C(Mean) | % bias | p-value | |
| Team size | 4.638 | 10.664 | -56.509 | 0.000 | 7.468 | 7.468 | 0.000 | 1.000 | 100 |
| International team | 0.275 | 0.399 | -30.929 | 0.000 | 0.327 | 0.327 | 0.000 | 1.000 | 100 |
| Number of references | 40.056 | 42.658 | -6.101 | 0.000 | 41.121 | 41.234 | -0.275 | 0.779 | 95.493 |
| Gender diversity | 0.720 | 0.961 | -25.029 | 0.000 | 0.973 | 0.973 | 0.000 | 1.000 | 100 |
| Publication year | 2012.725 | 2012.699 | 0.001 | 0.194 | 2013.024 | 2013.024 | 0.000 | 1.000 | 100 |
| Avg. academic age of TL(s) | 12.764 | 14.987 | -14.837 | 0.000 | 12.772 | 12.703 | 0.546 | 0.604 | 96.320 |
| Avg. prior productivity of TL(s) | 41.180 | 63.045 | -34.683 | 0.000 | 44.518 | 45.310 | -1.747 | 0.462 | 94.963 |

Table s3 The covariates balance check for the CEM3 model (T: RTL >0.7; C: RTL <=0.3)

| Variable | Before matching | | | | After matching | | | | %bias (reduction) |
|---|---|---|---|---|---|---|---|---|---|
| | T(Mean) | C(Mean) | % bias | p-value | T(Mean) | C(Mean) | % bias | p-value | |
| Team size | 4.986 | 9.872 | -49.497 | 0 | 7.264 | 7.264 | 0.000 | 1.000 | 100 |
| International team | 0.299 | 0.395 | -24.313 | 0 | 0.355 | 0.355 | 0.000 | 1.000 | 100 |
| Number of references | 40.423 | 42.849 | -5.663 | 0 | 41.792 | 41.740 | 0.125 | 0.843 | 97.793 |
| Gender diversity | 0.761 | 0.955 | -20.283 | 0 | 0.954 | 0.954 | 0.000 | 1.000 | 100 |
| Publication year | 2012.709 | 2012.773 | -0.003 | 0 | 2012.924 | 2012.924 | 0.000 | 1.000 | 100 |
| Avg. academic age of TL(s) | 13.054 | 14.780 | -11.678 | 0 | 13.338 | 13.272 | 0.499 | 0.449 | 95.727 |
| Avg. prior productivity of TL(s) | 42.992 | 62.511 | -31.225 | 0 | 47.203 | 47.679 | -0.998 | 0.484 | 96.804 |

Table s4 The covariates balance check for the CEM4 model (T: RTL >0.6; C: RTL <=0.4)

| Variable | Before matching | | | | After matching | | | | %bias (reduction) |
|---|---|---|---|---|---|---|---|---|---|
| | T(Mean) | C(Mean) | % bias | p-value | T(Mean) | C(Mean) | % bias | p-value | |
| Team size | 5.193 | 8.913 | -41.734 | 0 | 6.830 | 6.830 | 0.000 | 1.000 | 100 |
| International team | 0.306 | 0.380 | -19.539 | 0 | 0.344 | 0.344 | 0.000 | 1.000 | 100 |
| Number of references | 40.829 | 42.798 | -4.602 | 0 | 42.031 | 42.020 | 0.028 | 0.950 | 99.392 |
| Gender diversity | 0.774 | 0.935 | -17.247 | 0 | 0.927 | 0.927 | 0.000 | 1.000 | 100 |
| Publication year | 2012.681 | 2012.763 | -0.004 | 0 | 2012.862 | 2012.862 | 0.000 | 1.000 | 100 |
| Avg. academic age of TL(s) | 13.232 | 14.638 | -9.604 | 0 | 13.483 | 13.456 | 0.200 | 0.663 | 97.918 |
| Avg. prior productivity of TL(s) | 44.095 | 60.439 | -27.042 | 0 | 47.570 | 47.859 | -0.604 | 0.528 | 97.766 |

Table s5 The relationship between TL size and team performance

| Variable | (1)<br>Team impact | (2)<br>Team impact | (3)<br>Team impact | (4)<br>Team impact | (5)<br>Team disruptiveness | (6)<br>Team disruptiveness | (7)<br>Team disruptiveness | (8)<br>Team disruptiveness |
|---|---|---|---|---|---|---|---|---|
| RTL | -0.277***<br>(0.009) | | 0.028***<br>(0.010) | 0.204***<br>(0.040) | -2.806***<br>(0.295) | | -5.451***<br>(0.331) | -12.590***<br>(1.413) |
| RTL squared | | | | -0.146***<br>(0.032) | | | | 5.941***<br>(1.149) |
| Team size | | 0.031***<br>(0.001) | 0.032***<br>(0.001) | 0.033***<br>(0.001) | | -0.192***<br>(0.025 | -0.352***<br>(0.027) | -0.369***<br>(0.028) |
| Institutional diversity | | -0.001<br>(0.002) | -0.002<br>(0.002) | -0.002<br>(0.002) | | -0.044<br>(0.060) | 0.040<br>(0.060) | 0.038<br>(0.060) |
| International team | | 0.035***<br>(0.005) | 0.036***<br>(0.005) | 0.035***<br>(0.005) | | -3.172***<br>(0.187) | -3.250***<br>(0.187) | -3.206***<br>(0.187) |
| Number of references | | 0.009***<br>(0.000) | 0.009***<br>(0.000) | 0.009***<br>(0.000) | | 0.034***<br>(0.004) | 0.034***<br>(0.004) | 0.035***<br>(0.004) |
| Gender diversity | | 0.033***<br>(0.007) | 0.036***<br>(0.007) | 0.033***<br>(0.007) | | 0.622**<br>(0.249) | 0.145<br>(0.251) | 0.243<br>(0.251) |
| Avg. academic age of TL(s) | | -0.003***<br>(0.000) | -0.003***<br>(0.000) | -0.003***<br>(0.000) | | -0.267***<br>(0.013) | -0.270***<br>(0.013) | -0.269***<br>(0.013) |
| Avg. prior productivity of TL(s) | | 0.001***<br>(0.000) | 0.001***<br>(0.000) | 0.001***<br>(0.000) | | 0.011***<br>(0.002) | 0.010***<br>(0.002) | 0.010***<br>(0.002) |
| Constant | 2.819***<br>(0.005) | 2.395***<br>(0.019) | 2.373***<br>(0.021) | 2.332***<br>(0.023) | 51.661***<br>(0.180) | 42.029***<br>(0.632) | 46.281***<br>(0.682) | 47.957***<br>(0.753) |
| Field-fixed effect | No | Yes | Yes | Yes | No | Yes | Yes | Yes |
| Year-fixed effect | No | Yes | Yes | Yes | No | Yes | Yes | Yes |
| Observations | 127,023 | 127,023 | 127,023 | 127,023 | 127,023 | 127,023 | 127,023 | 127,023 |
| Adj R-squared | 0.0079 | 0.1472 | 0.1473 | 0.1474 | 0.0007 | 0.0416 | 0.0437 | 0.0439 |

Table s6 the causal effects of TL size on team performance

| Model | Groups | $\|T^{before}\|$ | $\|C^{before}\|$ | $\|T^{after}\|$ | $\|C^{after}\|$ | ATT (Number of citations) | $CI_{0.95}$ | P-value |
|---|---|---|---|---|---|---|---|---|
| CEM1 | T: RTL >0.9<br>C: RTL <=0.1 | 21643 | 1611 | 218 | 218 | 0.119 | [-3.940,4.179] | 0.954 |
| CEM2 | T: RTL>0.8<br>C: RTL <=0.2 | 24768 | 12603 | 3827 | 3827 | 0.853 | [-0.182,1.888] | 0.106 |
| CEM3 | T: RTL >0.7<br>C: RTL <=0.3 | 34986 | 28325 | 9497 | 9497 | 0.734 | [0.089,1.379] | 0.026 |
| CEM4 | T: RTL >0.6<br>C: RTL <=0.4 | 45776 | 49169 | 19271 | 19271 | 0.599 | [0.149,1.048] | 0 |
| CEM5 | T: RTL >0.5<br>C: RTL <=0.5 | 55795 | 71217 | 31681 | 31681 | 0.411 | [0.067,0.755] | 0.019 |

Table s7 The association between the number of thought leaders and team impact

| **Variables** | **Team impact** |
|---|---|
| RTL | 0.097***<br>(0.018) |
| RTL squared | -0.063***<br>(0.015) |
| Team size | 0.015***<br>(0.001) |
| Institutional diversity | 0.001<br>(0.001) |
| International team | 0.013***<br>(0.002) |
| Number of references | 0.004***<br>(0.000) |
| Gender diversity | 0.008***<br>(0.003) |
| Avg. academic age of TLs | -0.002***<br>(0.000) |
| Avg. prior productivity of TLs | 0.0004***<br>(0.000) |
| Constant | 0.270***<br>(0.011) |
| Field-fixed effect | Yes |
| Year-fixed effect | Yes |
| Observations | 141,550 |
| Adj R-squared | 0.0809 |

Notes: TL denotes thought leader; RTL denotes the ratio of thought leaders; Standard errors in parentheses. *** $p < 0.01$, ** $p < 0.05$, * $p < 0.1$.